\documentclass[aps,prd,twocolumn,groupedaddress,nofootinbib,preprintnumbers,superscriptaddress]{revtex4-2}  
\usepackage{graphicx,mathtools,tabu,array}
\usepackage{epstopdf}
\usepackage{amsmath}
\usepackage{amsfonts}
\usepackage[colorlinks=true,citecolor=red,linkcolor=blue,breaklinks=true]{hyperref}
\usepackage{accents}
\usepackage[figure, table]{hypcap}
\usepackage{amssymb}
\usepackage{appendix}
\usepackage{comment}
\usepackage{bbold}
\usepackage{xcolor}
\usepackage{slashed}
\usepackage{subfigure}
\usepackage{setspace}
\usepackage{footnote}
\usepackage{multirow}
\usepackage{braket}
\usepackage[normalem]{ulem}
\usepackage[utf8]{inputenc}
\usepackage{mathrsfs}
\usepackage{longtable}
\usepackage{enumitem}
\usepackage{orcidlink}

\newcommand{\CNB}{{\rm C}\nu {\rm B}}

\DeclareMathAlphabet{\mathpzc}{OT1}{pzc}{m}{it}

\graphicspath{{Figures/}}

\allowdisplaybreaks

\definecolor{palatd}{RGB}{104, 36, 109}
\definecolor{palatb}{RGB}{0, 56, 168}
\definecolor{palatr}{rgb}{0.745,0.118,0.176}
\newcommand\myshade{80}
\colorlet{mylinkcolor}{palatr}
\colorlet{mycitecolor}{palatb}
\colorlet{myurlcolor}{palatd}

\hypersetup{
  linkcolor  = mylinkcolor!\myshade!black,
  citecolor  = mycitecolor!\myshade!black,
  urlcolor   = myurlcolor!\myshade!black,
  colorlinks = true
}

\def\cnub{C$\nu$B~}
\def\dm{\delta m^2}

\def\eV{{\rm eV}}

\begin{document}
\sloppy  

\preprint{IPPP/23/37}

\title{From Dirac to Majorana: the Cosmic Neutrino Background capture rate in\\ the minimally extended Standard Model}

\author{Yuber F. Perez-Gonzalez\,\orcidlink{0000-0002-2020-7223}}
\email{yuber.f.perez-gonzalez@durham.ac.uk}
\affiliation{Institute for Particle Physics Phenomenology, Durham University, South Road DH13EL, Durham, United Kingdom}

\author{Manibrata Sen\,\orcidlink{0000-0001-7948-4332}}
\email{manibrata@mpi-hd.mpg.de}
\affiliation{Max-Planck-Institut f\"ur Kernphysik, Saupfercheckweg 1, 69117 Heidelberg, Germany}

\begin{abstract}
  We investigate the capture rate of the cosmic neutrino background on tritium within the Standard Model, extended to incorporate three right-handed singlet neutrinos with explicit lepton-number violation. 
  We consider a scenario where the $6 \times 6$ neutrino mixing matrix factorizes into three independent $2 \times 2$ pairs and analyze the states produced from weak interactions just before neutrino decoupling. 
  Taking into account the unrestricted Majorana mass scale associated with lepton number violation, spanning from the Grand Unification scale to Planck-suppressed values, we observe a gradual transition in the capture rate from a purely Majorana neutrino to a purely (pseudo) Dirac neutrino. We demonstrate that the capture rate is modified if the lightest active neutrino is relativistic, and this can be used to constrain the tiniest value of mass-squared difference $\sim 10^{-35}\,{\rm eV}^2$, between the active-sterile pair, probed so far. Consequently, the cosmic neutrino capture rate could become a promising probe for discerning the underlying mechanism responsible for generating neutrino masses.
\end{abstract}

\maketitle

\section{Introduction}

Standard cosmology predicts that the present Universe is awash with a sea of neutrinos, produced approximately a second after the Big Bang. This Cosmic Neutrino Background ($\CNB$) is a sea of relic neutrinos, much like the Cosmic Microwave Background (CMB) is a sea of relic photons left after photon decoupling around 380,000 years after the Big Bang~\cite{Dicke:1965zz}. 
Since the \cnub is much older than the CMB, a careful study of the \cnub is crucial for a better understanding of the early Universe. 

The neutrinos composing the \cnub are expected to follow a Fermi-Dirac distribution\footnote{This is true in the absence of neutrino clustering~\cite{Singh:2002de, Ringwald:2004np}, an assumption we make in this work}, with a temperature today of around $1.95\,{\rm K}$, which is $(4/11)^{1/3}$ the temperature of the CMB photons today. This happened due to the temperature of the photons increasing during electron-positron decoupling at around $0.5\,{\rm MeV}$. The neutrinos, on the other hand, decoupled from the plasma at around $1\,{\rm MeV}$. For the present day CMB temperature $T_{\gamma 0}=0.23\,{\rm meV}$, the present day neutrino temperature $T_{\nu 0}=0.17\,{\rm meV}$. 
Thus, following a Fermi-Dirac distribution, the current neutrino number density today is $\sim 112\,{\rm cm}^{-3}$ per flavor. 
The helicity distribution of this neutrino number density depends on the neutrino nature.
For Dirac neutrinos, we expect that only left-helical neutrinos and right-helical antineutrino states are populated, while for Majorana, both left- and right-helical states should be present in the $\CNB$.
Furthermore, from the bounds on neutrino masses from neutrino oscillation experiments, $m_{\nu_2}\geq \sqrt{\Delta m^2_{\rm sol}}= 8.7\,{\rm meV}$, $m_{\nu_3}\geq \sqrt{\Delta m^2_{\rm atm}}=48\,{\rm meV}$ in normal mass ordering, and $m_{\nu_2}\geq \sqrt{\Delta m^2_{\rm atm}}=48\,{\rm meV}$, $m_{\nu_1} \geq \sqrt{\Delta m^2_{\rm atm}-\Delta m^2_{\rm sol}}=47\,{\rm meV}$, we have that at least two of the neutrinos will be non-relativistic today~\cite{Esteban:2020cvm}.

An experimental detection of the \cnub will not only present us with a validation of our understanding of the early Universe but also present the first-ever detection of non-relativistic neutrinos. As a result, a lot of theoretical as well as experimental efforts are underway to detect the \cnub. Currently, the most popular and feasible idea is that of neutrino capture on beta-decaying nuclei, postulated first by Weinberg~\cite{Weinberg:1962zza}. The PTOLEMY experiment~\cite{PTOLEMY:2018jst} aims at detecting the \cnub through neutrino capture on tritium: $\nu + {}^3{\rm H} \rightarrow {}^3{\rm He}^+ + {\rm e}^- $. The signal at PTOLEMY will be an electron emitted with kinetic energy equalling $2\,m_\nu$ above the beta decay endpoint. Nevertheless, there are a number of experimental and theoretical challenges, in particular, with attaining an energy resolution as low as $0.1\,{\rm eV}$ with current technology. This is currently an open issue and a lot of experimental and technological efforts are underway to overcome this barrier~\cite{Cheipesh:2021fmg,Mikulenko:2021ydo,Cheipesh:2023qiy,PTOLEMY:2022ldz}. Apart from this, a number of other ideas has been proposed to detect the \cnub ~\cite{Stodolsky:1974aq,Shvartsman:1982sn,Akhmedov:2019oxm, Chao:2021ahl, Shergold:2021evs,Bauer:2021uyj,Brdar:2022wuv}. However, these are futuristic and cannot be achieved in the near foreseeable future. The capture rate also depends quite sensitively on whether the \cnub clusters or not~\cite{Singh:2002de, Ringwald:2004np, Arvanitaki:2022oby}. A comprehensive discussion of the different constraints on neutrino clustering is given in~\cite{Bauer:2022lri}.

A direct detection of the \cnub will be crucial to testing fundamental properties associated with neutrinos such as their lifetime, whether they cluster or not~\cite{Ringwald:2004np,Mertsch:2019qjv,Brdar:2022kpu}, additional interactions of neutrinos~\cite{Arteaga:2017zxg,Akita:2021hqn, Alvey:2021xmq, Das:2022xsz, Banerjee:2023lrk} and so on. These neutrinos, being non-relativistic, will allow us to probe kinematical regions, which are otherwise inaccessible in terrestrial laboratories. For example, detecting the \cnub can be used to differentiate between the Dirac and Majorana nature of neutrinos~\cite{Long:2014zva, Roulet:2018fyh}. If the neutrinos are Majorana particles, then the capture rate will be two times more than that for Dirac neutrinos when all three mass eigenstates are non-relativistic today (see text for more details). This can act as a direct test for lepton number violation in the Standard Model (SM).

However, it is possible that lepton number is violated \emph{softly} in the SM. The extent of lepton number violation (LNV) can be quantified through the smallness of the Majorana mass term, in comparison to the Dirac mass term for neutrinos. In such a scenario, neutrinos are pseudo-Dirac (or quasi-Dirac)~\cite{Wolfenstein:1981kw,Petcov:1982ya,Bilenky:1983wt,Kobayashi:2000md,Anamiati:2017rxw,deGouvea:2009fp,Vissani:2015pss}. The softness of LNV guarantees that although neutrinos are Majorana in nature, they behave as Dirac neutrinos for all practical purposes. Active-sterile neutrino oscillations are usually driven by a tiny mass-squared difference $(\dm)$ between the mass-eigenstates and could be accessible only over astronomically large baselines. Strong constraints on the smallness of the mass-squared difference arise from high-energy neutrinos, $10^{-18}~\eV^2\lesssim \dm \lesssim 10^{-12}~\eV^2$~\cite{Rink:2022nvw, Carloni:2022cqz}, supernova neutrinos~$\dm\lesssim 10^{-20}~\eV^2$~\cite{deGouvea:2020eqq,Martinez-Soler:2021unz,Sen:2022mun} as well as solar neutrinos~$\dm\lesssim 10^{-11}~\eV^2$~\cite{deGouvea:2009fp,Ansarifard:2022kvy, Franklin:2023diy}. Weaker constraints also exist from neutrino oscillation experiments~\cite{Das:2014jxa,Hernandez:2018cgc,Anamiati:2019maf} as well as atmospheric neutrinos, $\dm\lesssim 10^{-4}~\eV^2$~\cite{Beacom:2003eu}. 

\begin{figure}[!t]
\centering
\includegraphics[width=\linewidth]{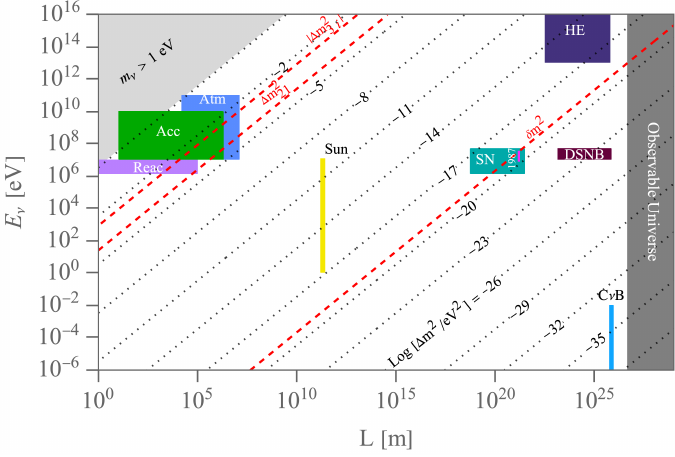}
\caption{Landscape of neutrino mass-squared difference $\dm$ in the neutrino energy $(E_\nu)$ and experiment baseline $(L)$ plane. The corresponding sensitivity from reactor neutrinos (light purple), accelerator neutrinos (green), atmospheric neutrinos (blue), solar neutrinos (yellow), supernova neutrinos (emerald), diffuse supernova neutrino background (dark red) and high energy neutrinos (purple) are shown. The bound from neutrino data from SN1987A is shown by a pink region. Predictions from the \cnub derived in this work, assuming the lightest neutrino to be relativistic today, are shown in light blue. The dashed red lines indicate the solar, atmospheric mass splittings $\Delta m_{21}^2, |\Delta m_{3i}^2|$ and a value of $\delta m = 6.31\times 10^{-20}~{\rm eV^2}$ preferred by the SN1987A data~\cite{Martinez-Soler:2021unz}.}
\label{fig:LvE}
\end{figure}

If neutrinos are pseudo-Dirac, it would also affect the cosmic neutrino capture rate. One would expect there to be a gradual transition from the capture rate in the Dirac case to that in the Majorana case, and this transition should be a function of the extent of LNV, given by $\dm$. Therefore, when $\dm$ is tiny, we expect the capture rate to behave like that for Dirac neutrinos. On the other hand, for large $\dm$, we should recover the Majorana capture rate. Furthermore, the rate is also modified if the lightest neutrino is relativistic at the time of capture, thereby allowing a probe of the smallness of $\dm$. These differences in capture rate would clearly show up in an experiment like PTOLEMY, thereby allowing a complementary probe of LNV through the \cnub. We show that PTOLEMY will be sensitive to $\dm \sim 10^{-35}\,{\rm eV}^2$ - easily shadowing the sensitivity from all other sources of LNV, and therefore set the strongest constraints on the smallness of $\dm$. This is demonstrated in Fig.\,\ref{fig:LvE}, which shows the sensitivity of different neutrino sources to $\dm$ in the $E_\nu-L$ plane. Clearly, positive detection of the \cnub can be used to constrain the tiniest value of $\dm$ probed so far.

The paper is organised as follows. In Sec.\,\ref{sec:model}, we discuss the minimally extended Standard Model, by adding 3 singlet neutrinos and explore the mass-squared differences between the active-sterile neutrinos. In Sec.\,\ref{sec:capture}, we discuss the capture rate of the cosmic neutrino background in the case of soft violation of lepton number. In Sec.\,\ref{sec:Ptolemy}, we demonstrate the event rates in an upcoming experiment like PTOLEMY. Finally, we conclude in Sec.\,\ref{sec:conc}.
We consider natural units where $\hbar = c = k_{\rm B} = 1$ throughout this manuscript.

\section{A minimal Standard Model Extension}
\label{sec:model}

The gauge symmetries of the Standard Model (SM) allow for the existence of singlets with zero hypercharges, which can couple to the left-handed lepton doublets and generate Yukawa terms responsible for neutrino masses. Initially, one might expect these Yukawa couplings to be extremely small, of the order of ${\cal O}(10^{-12})$, in order to match the observed neutrino mass scale of ${\cal O}({\rm eV})$.
However, it is worth noting that the same SM symmetries also permit Majorana mass terms for those singlets.
While such terms lead to lepton number violation, this symmetry is accidental and does not pose any fundamental issues. 
Furthermore, the scale of these Majorana mass terms is only loosely constrained~\cite{deGouvea:2009fp}.
In fact, it can be close to the scale of Grand Unification Theories (GUT), or it can be suppressed relative to the electroweak scale.
In the first case, corresponding to the well-known see-saw mechanism, the Majorana mass terms are at the GUT scale. 
In the second case, known as the Pseudo-Dirac scenario, the mass terms are suppressed compared to the electroweak scale. 
Let us examine these scenarios in greater detail.
The mass Lagrangian for neutrinos, which includes both Yukawa interactions with the singlets $\nu_{R}^i$, $i=\{1,2,3\}$, and their Majorana mass terms, can be written as
\begin{align}
    \mathscr{L}_{\nu}  = -Y_{\alpha i} \overline{L_\alpha} \widetilde{H} \nu_{R}^i + \frac{1}{2}\overline{(\nu_{R}^i)^c} M_R^{ij} \nu_{R}^j\,.
\end{align}
Here, $Y_{\alpha i}$ represents the Yukawa couplings between the left-handed lepton doublets $L_\alpha$, the conjugate of the SM Higgs doublet $\widetilde{H}$, and the singlets. The Majorana mass term, denoted by $M_R^{ij}$, depends on the scale at which such terms originate. The superscript $c$ signifies charge conjugation.
After electroweak symmetry breaking, the neutrino mass Lagrangian can be rewritten as
\begin{align}
\mathscr{L}_{\nu} = - \frac{1}{2} \overline{N_L^c} M N_L,
\end{align}
where
\begin{align}
N_L = \begin{pmatrix}
\nu_L\\
(\nu_R)^c
\end{pmatrix},\quad
M = \begin{pmatrix}
0_3 & Y v/\sqrt{2}\\
Y v/\sqrt{2} & M_R
\end{pmatrix}.
\end{align}
In the above expressions, $v$ represents the vacuum expectation value (VEV) of the Higgs field, $\nu_L = (\nu_e, \nu_\mu, \nu_\tau)^T$ denotes the left-handed neutrino field, and $\nu_R=(\nu_{R_1}, \nu_{R_2}, \ldots)^T$ represents the right-handed neutrino field. At this stage, we have not specified any hierarchy between the Higgs VEV and the scale of the Majorana mass matrix $M_R$.

In scenarios where a significant hierarchy exists between the Majorana mass and the electroweak scales, i.e., $M_R \gg Yv$, the diagonalization of the matrix $M$ gives rise to active neutrinos with suppressed masses relative to the electroweak scale, $m_\nu \propto Y^T (M_R)^{-1} Y v^2$.
This mechanism, widely known as the seesaw mechanism~\cite{Mohapatra:1979ia, Gell-Mann:1979vob, Yanagida:1979as, Minkowski:1977sc, Mohapatra:1980yp, Magg:1980ut, Lazarides:1980nt, Wetterich:1981bx, Foot:1988aq, Ma:1998dn}, has garnered considerable attention due to its potential to explain the observed matter-antimatter asymmetry in the Universe~\cite{Yanagida:1979as}.

However, it is also plausible that the Majorana mass scale is suppressed relative to the electroweak scale, $M_R \ll Yv$, particularly if the Majorana mass terms are Planck-suppressed, for example. 
In this particular scenario, referred to as the ``pseudo-Dirac'' case, lepton number is softly broken by the Majorana mass, resulting in the lifting of degeneracy between the left- and right-handed components of a Dirac neutrino.
Significantly, in this scenario, processes involving lepton-number violation are highly suppressed, making it challenging to experimentally detect lepton-number violating phenomena.

In order to test the pseudo-Dirac scenario, it is then crucial to explore the consequences of the presence of Majorana mass terms, particularly for the oscillations between the active and sterile neutrino components. 
Let's first consider the general case where we do not assume any specific hierarchy between the Majorana mass matrix and the electroweak scale. 
The mass matrix $M$ can be diagonalized by a $6\times 6$ unitary matrix, $\mathscr{V}$, which is obtained from the multiplication of 15 complex rotation matrices~\cite{Anamiati:2019maf}. 
For simplicity, we will focus on the mixing between the pseudo-Dirac pairs labelled as $1-4$, $2-5$, and $3-6$.
Hence, considering only as non-zero mixing angles $\theta_{14}, \theta_{25}, \theta_{36}$, the mixing matrix $\mathscr{V}$ can be expressed as
\begin{align}
    \mathscr{V} = U_{23} U_{13} U_{12} U_{14} U_{25} U_{36}\,.
\end{align}
We therefore define the mass eigenstates $\nu_{i}^\pm$~\cite{Kobayashi:2000md}
\begin{align*}
    N_L = \mathscr{V}
            \begin{pmatrix}
                \nu_{i}^-\\
                \nu_{i}^+
            \end{pmatrix}\,,
\end{align*}
where $\pm$ refers to the two mass eigenstates associated with the splitting of a given mass eigenstate $i$.
Assuming the singlet mass matrix $M_R$ to be diagonal, $M_R={\rm diag}(m_{r_1}, m_{r_2}, m_{r_3})$, we have the masses $m_{i}^\pm$ associated to the eigenstates
\begin{align}
    m_{i}^\pm = \frac{1}{2}\left[\sqrt{(m_{r_i})^2+(2 m_{D_i})^2} \pm m_{r_i}\right]\,,
\end{align}
with $m_{D_i}= Y v/\sqrt{2}$ being the eigenvalues of the Dirac mass matrix.
Therefore, the mixing angle for each generation will be
\begin{align}
    \tan 2\theta_{i} = \frac{2m_{D_i}}{m_{r_i}}\,.
\end{align}
In our case, where only mixing between the pseudo-Dirac pairs $1-4$, $2-5$, and $3-6$ are considered, this implies $\theta_{1,2,3}=\theta_{14,25,36}$. 
Explicitly, the neutrino fields in the flavor basis take a simple form
\begin{align}\label{eq:nu_active}
    \nu_{\alpha} = \sum_{i}\,U_{\alpha i} (e^{\mathrm{i}\lambda} \cos\theta_{i}\, \nu_i^- + \sin\theta_{i}\,\nu_i^+)\,,
\end{align}
with $U_{\alpha i}$ the standard Pontecorvo-Maki-Nakagawa-Sakata mixing matrix.
We observe that a flavor eigenstate corresponds to a superposition of six mass eigenstates $\nu_{i}^\pm$.
The CP phase $e^{\mathrm{i}\lambda}$ in Eq.~\eqref{eq:nu_active} is fixed after imposing the masses to be positive, finding that $e^{\mathrm{i}\lambda} = \mathrm{i}$~\cite{GiuntiKim}.
The orthogonal components $\nu_{\{s_1,s_2,s_3\}}$, which represent the states that do not interact weakly, can be written as
\begin{align}\label{eq:sterf}
    \nu_{s_i} &= -\mathrm{i} \sin\theta_{i}\, \nu_i^- + \cos\theta_{i}\,\nu_i^+.
\end{align}

Let us now consider in detail the limits mentioned before of this scenario depending on the scale of the singlet mass matrix $M_R$.\\
\emph{See-saw limit: $M_R\gg Yv$.} In such a case, we have that the mixing becomes tiny, $\theta_{i}\to 0$, in such a way that the flavor and sterile fields become,
\begin{align}\label{eq:ss}
    \nu_{\alpha} &\approx \mathrm{i}\sum_i\, U_{\alpha i} \nu_i^-, \quad \nu_{s_i} \approx \nu_i^+,
\end{align}
such that the states $\nu_i^\pm$ have masses
\begin{align}
    m_i^- = \frac{(m_{D_i})^2}{m_{r_i}}, \quad m_i^+ = m_{r_i}\,.
\end{align}
This indicates that sterile neutrinos are mostly composed of $\nu_i^+$ eigenstates, while flavor states are superpositions of the $\nu_i^-$ states, which we can identify as the usual mass eigenstate fields.\\
\emph{Pseudo-Dirac limit: $M_R\ll Yv$.} In such a case, we have that the mixing becomes maximal, $\theta_{i}\to \pi/4$, and the flavor and sterile fields become,
\begin{subequations}
   \begin{align}\label{eq:ss}
        \nu_{\alpha} &= \sum_i \frac{U_{\alpha i}}{\sqrt{2}}(\,\mathrm{i}\,\nu_i^- + \nu_i^+),\\
        \nu_{s_i} &= \frac{1}{\sqrt{2}}(\,-\mathrm{i}\, \nu_i^- + \nu_i^+).
    \end{align} 
\end{subequations}
Here the masses for the mass eigenstates are given by
\begin{align}
    m_i^\pm = m_{D_i} \pm \frac{m_{r_i}}{2},
\end{align}
respectively. 
Note that when we consider the exact Dirac case, $m_{r_i}=0$, we recover the usual fact that a neutral Dirac field is a maximally mixed superposition of two degenerate Majorana neutrinos.

Now, to establish the specific properties of the relic neutrinos in the PD scenario, we have to first determine the states participating in the weak interactions, a task which will be considered in the next subsection.

\subsection{Weak Interactions}
Before their decoupling, neutrinos were in an ultra-relativistic state and in thermal equilibrium due to their weak interactions. 
As the Universe cooled down, neutrinos decoupled from the thermal bath, and will therefore retain the flavor state related to their last scattering.
Thus, the initial states will be linear superpositions of the mass eigenstates $\nu_i^\pm$. 
However, since weak interactions violate parity, it becomes crucial to carefully determine the specific superposition that is emitted based on the weak process involved. 
In simpler terms, we need to specify whether the initial state created has a right or left helicity.
To address this, we can examine the charged-current (CC) weak interaction Lagrangian explicitly, which is written using the defined flavor fields mentioned above,
\begin{widetext}
    \begin{align}\label{eq:CCL}
        \mathscr{L}_{\rm CC}&=-\frac{g}{\sqrt{2}}\sum_{\alpha=e,\mu,\tau}[\overline{\nu_\alpha}\gamma^\mu\alpha_L W_\mu + \overline{\alpha_L}\gamma^\mu\nu_\alpha W_\mu^\dagger],\notag\\
        &=-\frac{g}{\sqrt{2}}\sum_{\alpha=e,\mu,\tau}\sum_{i}[U_{\alpha i}^*(-\mathrm{i} \cos\theta_{i}\, \overline{\nu_i^-} + \sin\theta_{i}\,\overline{\nu_i^+})\gamma^\mu\alpha_L W_\mu + U_{\alpha i}\overline{\alpha_L}\gamma^\mu(\mathrm{i} \cos\theta_{i}\, \nu_i^- + \sin\theta_{i}\,\nu_i^+) W_\mu^\dagger].
    \end{align}
Examining this Lagrangian, we notice that the two currents yield distinct linear combinations. 
To determine the helicities of these combinations, let's recall the expansion of a generic Majorana field operator $\psi$,
\begin{align}
\psi(x)=\int\frac{d^3p}{(2\pi)^32E} \sum_{h=\pm}[a_h(p) u_h(p)e^{-\mathrm{i}px}+a_h^\dag(p) v_h(p)e^{\mathrm{i}px}],
\end{align}
where, $u_\pm$ and $v_\pm$ represent four-component spinors, and $a$ and $a^\dagger$ are quantum operators adhering to standard anticommutation relations. 
It follows that the operator $\psi$ can create or annihilate the same state, as expected from a Majorana fermion.
Given that neutrinos were ultra-relativistic at decoupling, we can consider the following approximations for the spinors $u_\pm$ and $v_\pm$~\cite{GiuntiKim}
\begin{align}
    u_+(p)&\approx -\sqrt{2E}\begin{pmatrix}
    \chi^+(p)\\
    -\frac{m}{2E}\chi^+(p)
    \end{pmatrix},&
    u_-(p)&\approx \sqrt{2E}\begin{pmatrix}
    -\frac{m}{2E}\chi^-(p)\\
    \chi^-(p)
    \end{pmatrix}\notag\\
    v_+(p)&\approx -\sqrt{2E}\begin{pmatrix}
    \frac{m}{2E}\chi^-(p)\\
    \chi^-(p)
    \end{pmatrix},&
    v_-(p)&\approx \sqrt{2E}\begin{pmatrix}
    \chi^+(p)\\
    \frac{m}{2E}\chi^+(p)
    \end{pmatrix},
\end{align}
\end{widetext}
where $\chi^\pm$ are two-component helicity eigenstate spinors.

Hence, the first terms of the charged-current (CC) Lagrangian in Eq.~\eqref{eq:CCL}, $\overline{\nu_i^\pm}\gamma^\mu\alpha_L W_\mu$, create a $\nu_i^\pm$ with negative helicity ($h=-1$ or a \emph{neutrino}) or annihilate a $\nu_i^\pm$ with positive helicity ($h=+1$ or an \emph{antineutrino}). 
The second term operates conversely, creating neutrinos with positive helicity and annihilating neutrinos with negative helicity.
Thus, the neutrino states with negative helicity, $|\nu_\alpha\rangle_{h=-1}$, and positive helicity, $|\overline{\nu}_\alpha\rangle_{h=1}$, created by the CC Lagrangian correspond to the following linear superpositions,
\begin{subequations}
    \begin{align}\label{eq:stat}
        |\nu_\alpha\rangle_{h=-1} &= U_{\alpha i}^* (-\mathrm{i} \cos\theta_{i}|\nu_i^-\rangle + \sin\theta_{i}|\nu_i^+\rangle)\\
        |\overline{\nu}_\alpha\rangle_{h=1} &= U_{\alpha i} ( \mathrm{i} \cos\theta_{i}|\nu_i^-\rangle + \sin\theta_{i}|\nu_i^+\rangle) .
    \end{align}
\end{subequations}
The conjugation arises from the nature of the interaction entering the CC Lagrangian. 
In the previously described see-saw limit, the states $|\nu_\alpha\rangle_{h=-1}$ and $|\overline{\nu}_\alpha\rangle_{h=+1}$ take the approximate forms:
\begin{align*}
|\nu_\alpha\rangle_{h=-1} &\approx -\mathrm{i}\, U_{\alpha i}^* |\nu_i^-\rangle \\
|\overline{\nu}_\alpha\rangle_{h=1} & \approx \quad\! \mathrm{i}\, U_{\alpha i}|\nu_i^-\rangle,
\end{align*}
These expressions, up to an irrelevant overall phase $\pm \mathrm{i}$, align with the conventional definitions of neutrino and antineutrino states commonly employed in neutrino oscillation studies~\cite{GiuntiKim}.
In contrast, in the Dirac limit, the states are approximately given by:
\begin{align*}
|\nu_\alpha\rangle_{h=-1} &\approx \frac{U_{\alpha i}^*}{\sqrt{2}} (-\mathrm{i} |\nu_i^-\rangle + |\nu_i^+\rangle) \equiv U_{\alpha i}^* |\nu_i\rangle \\
|\overline{\nu}_\alpha\rangle_{h=1}&\approx \frac{U_{\alpha i}}{\sqrt{2}} ( \mathrm{i} |\nu_i^-\rangle + |\nu_i^+\rangle) \equiv U_{\alpha i}|\overline{\nu}_i\rangle,
\end{align*}
Again, these approximations are consistent with the standard mixing of neutrinos and antineutrinos after defining the neutrino mass eigenstate to be $|\nu_i\rangle = \frac{1}{2}(-\mathrm{i} |\nu_i^-\rangle + |\nu_i^+\rangle)$, while the antineutrino state is related by complex conjugation.
Thus, it is evident that the general superpositions defined in Eqs.~\eqref{eq:stat} correctly reproduce the expected limits for both the see-saw and Dirac scenarios.
As for the sterile state, it follows from Eq.~\eqref{eq:sterf}:
\begin{equation}
|\nu_{s_i}\rangle = - \mathrm{i} \sin\theta_{i}|\nu_i^-\rangle + \cos\theta_{i}|\nu_i^+\rangle.
\end{equation}
These are the potential superpositions in which neutrinos, both left- and right-handed would have frozen out after the decoupling phase. 
Moreover, as the mass eigenstates $|\nu_i^\pm\rangle$ evolve with distinct phases, there is a possibility that the initial flavor states would oscillate to sterile ones, which do not interact and would result in the disappearance of a portion of the \cnub. 
The occurrence of active-sterile oscillations is closely linked to the value of the neutrino capture rate for Dirac neutrinos, as we will explore in the following section.

\section{Capture rate computation}
\label{sec:capture}

Due to the non-relativistic nature of the neutrinos today, chirality and helicity can no longer be used interchangeably. We will work with helicities here. 
The tiny mass-squared difference between $\nu_i^\pm$ in the pseudo-Dirac scenario will induce active sterile oscillations, which can take place over baselines $\propto E/\delta m^2$. 
These oscillations conserve helicity, leading to $\nu^\alpha_{h=1}\longleftrightarrow\nu^s_{h=1}$ and $\nu^\alpha_{h=-1}\longleftrightarrow \nu^s_{h=-1}$.
Henceforth, we will drop the subscript $h$ and use $\pm 1$ to denote the helicity state of the neutrino.
Since relic neutrinos have propagated in an expanding Universe, the evolution phases from the decoupling, occurring at a redshift $z$, until today depending on the momentum $p$ are given by~\cite{Beacom:2004yd,Esmaili:2012ac},
\begin{align}
    \Phi_i^{\pm}(z)=\int_0^{z} \frac{dz^\prime}{H(z^\prime)} \left[(m_{i}^{\pm})^2 + p^2 (1+z^\prime)^2\right]^{\frac{1}{2}},
\end{align}
where $H(z) = H_0(1+z)\sqrt{\Omega_m (1+z)^3+\Omega_r(1+z)^4+\Omega_\Lambda}$ is the Hubble function, depending on the Hubble parameter $H_0$, and the matter, $\Omega_m$, radiation $\Omega_r$, and Dark Energy $\Omega_\Lambda$ contributions to the total energy density~\cite{Planck:2018vyg}.
Thus, the positive and negative helicity states will evolve according to
\begin{align*}
    |\nu_i^\pm (z) \rangle= \exp(-i\Phi^\pm(z)) |\nu_i^\pm \rangle
\end{align*}
The disappearance probability $P(\nu^\alpha_{\pm 1}\to \nu^{s_i}_{\pm 1})$ for each eigenstate $i$ is then
\begin{subequations}\label{eq:convprob}
    \begin{align}
        P(\nu^\alpha_{-1}\to \nu^{s_i}_{-1})&=|\langle \nu_{s_i}|\nu_\alpha(z) \rangle|^2\notag\\
        &= |U_{\alpha i}|^2\sin^2\,2\theta_{i} \sin^2\left[\frac{\Delta \Phi_i}{2}\right]\,,\\
        P(\nu^\alpha_{+1}\to \nu^{s_i}_{ +1})&=|\langle \nu_{s_i} |\overline{\nu}_\alpha(z) \rangle|^2 \notag\\
        &= |U_{\alpha i}|^2\sin^2\,2\theta_{i} \cos^2\left[\frac{\Delta \Phi_i}{2}\right]\,,
    \end{align}
\end{subequations}
where the phase difference is $\Delta \Phi_i = \Phi_i^+-\Phi_i^-$. 

After freeze-out, the phase-space distribution of the C$\nu$B remains a Fermi-Dirac distribution, while the temperature and the momenta redshift. 
Therefore, the abundance at freeze-out for effectively massless neutrinos is given as 
\begin{equation}
    n(T)=  \frac{3\zeta(3)}{4\pi^2}T_\nu^3\,.
\end{equation}
where $T_\nu$ is related to the photon temperature $T_\gamma$ through $T_\nu =(4/11)^{1/3} T_\gamma$. 
The current number density of neutrinos, after accounting for redshift, is $n_0\equiv 56\,{\rm cm}^{-3}$ per flavor per helicity state of the neutrino.
Moreover, we have that the root mean square momentum of neutrinos is $\overline{p}=0.6$ meV~\cite{Long:2014zva}, indicating that the two heaviest states are non-relativistic today, while the lightest could be still relativistic if it has a mass smaller than $\sim 0.1$ meV.
Since only the states $|\nu_\alpha\rangle_{-1}, |\overline{\nu}_\alpha\rangle_{+1}$ are populated in the Early Universe, in equal amounts, their abundances at present follow~\cite{Long:2014zva}
\begin{subequations}
\begin{align}
    n(\nu^\alpha_{-1}) &=n_0\,,\qquad\qquad n(\nu^\alpha_{+1}) = n_0\,,\\
    n(\nu^{s_i}_{\pm 1}) &= 0 ,
\end{align}
\end{subequations}
Note that we have assumed that the sterile states are not populated in the Early Universe.

Now, let us consider the effect of mixing between active and sterile states in our scenario. In this regime, after the neutrinos have decoupled, the sterile states can be populated with a probability given by Eq.\,\ref{eq:convprob}. As a result,  the abundances of the neutrinos are given by  
\begin{widetext}
\begin{align}
    n(\nu^\alpha_{-1}) &=(1-P(\nu^\alpha_{-1}\to \nu^{s_i}_{-1}))\,n_0\,,\qquad\qquad n(\nu^\alpha_{+1}) = (1-P(\nu^\alpha_{+1}\to \nu^{s_i}_{+1}))\,n_0\,,\\
    n(\nu^{s_i}_{-1}) &= P(\nu^\alpha_{i, -1}\to \nu^{s_i}_{-1}) \,n_0\,,\qquad\qquad\qquad\,\,\, n(\nu^{s_i}_{+1}) = P(\nu^\alpha_{+1}\to \nu^{s_i}_{+1})\, n_0\,.
\end{align}
Some of the active neutrinos will be lost from the thermal plasma due to active-sterile conversion. This is the main effect of having lepton number violation, after the addition of singlet states having Majorana masses. The Majorana limit can be recovered for $P(\nu^\alpha_{\pm 1}\to \nu^{s_i}_{\pm 1})=0$.

Taking into account the context discussed in this thread, let us now delve into the calculation of the capture rate of the \cnub on a target nuclei, represented by the process $\nu_e + n \to p^+ + e^-$. Following the established standard procedure to compute this rate~\cite{Long:2014zva}, we arrive at the following result
\begin{equation}
    \Gamma_{\CNB}= N_T\,\overline{\sigma}\,\sum_{i=1}^3 \left[(1-P(\nu^e_{+ 1}\to \nu^{s_i}_{+1})) \, n(\nu^\alpha_{+1})\,{\cal A}_i(+1)+(1-P(\nu^e_{-1}\to \nu^{s_i}_{-1}))\, n(\nu^\alpha_{-1})\,{\cal A}_i(-1)\right]\,,
\end{equation}
where $N_T$ are the number of targets, and ${\cal A}(h)$ are spin-dependent factors that take into account the mismatch between helicity and chirality,
\begin{align}
    {\cal A}_i(h)\equiv 1-h \overline{v_i},
\end{align}
being $\overline{v_i}=(v_i^++v_i^-)/2$, with $v_i^\pm=|\vec{p}|/\sqrt{|\vec{p}|^2+(m_i^\pm)^2}$ the average neutrino velocity, $h$ the helicity. The nucleus-dependent factor $\overline{\sigma}$ in the capture rate is the spin-averaged cross-section. Assuming tritium as the target, we have that 
\begin{align}
    \overline{\sigma} \approx 3.8\times 10^{-45}{\rm\ cm^2}.
\end{align}
Expanding the capture rate, we find the following dependence on the mixing between the $\nu_i^\pm$ fields,
\begin{align}\label{eq:CR}
    \Gamma_{\CNB} &= N_T\overline{\sigma}  n_0\sum_{i=1}^3 |U_{ei}|^2\left[1 + \cos^22\theta_{i} +  \sin^22\theta_{i}\,\langle v_i\cos\left(\Delta \Phi_i\right)\rangle \right],
\end{align}
\end{widetext}
where we have taken the average of the oscillatory term with respect to the $\CNB$ momentum distribution $f_{\CNB}(p)$~\cite{Roulet:2018fyh},
\begin{align}
    \langle v_i\cos\left(\Delta \Phi_i\right)\rangle = \frac{\int_0^\infty v_i\cos\left(\Delta \Phi_i\right) p^2 f_{\CNB}(p)\, dp}{\int_0^\infty p^2 f_{\CNB}(p)\, dp}.
\end{align}
As mentioned before, we consider a Fermi-Dirac distribution for the $\CNB$ momentum in terms of the temperature of the relic neutrinos today,
\begin{align*}
    f_{\CNB}(p) = \frac{1}{\exp(p/T_\nu) + 1}.
\end{align*}
\begin{figure*}[!t]
\centering
\includegraphics[width=\linewidth]{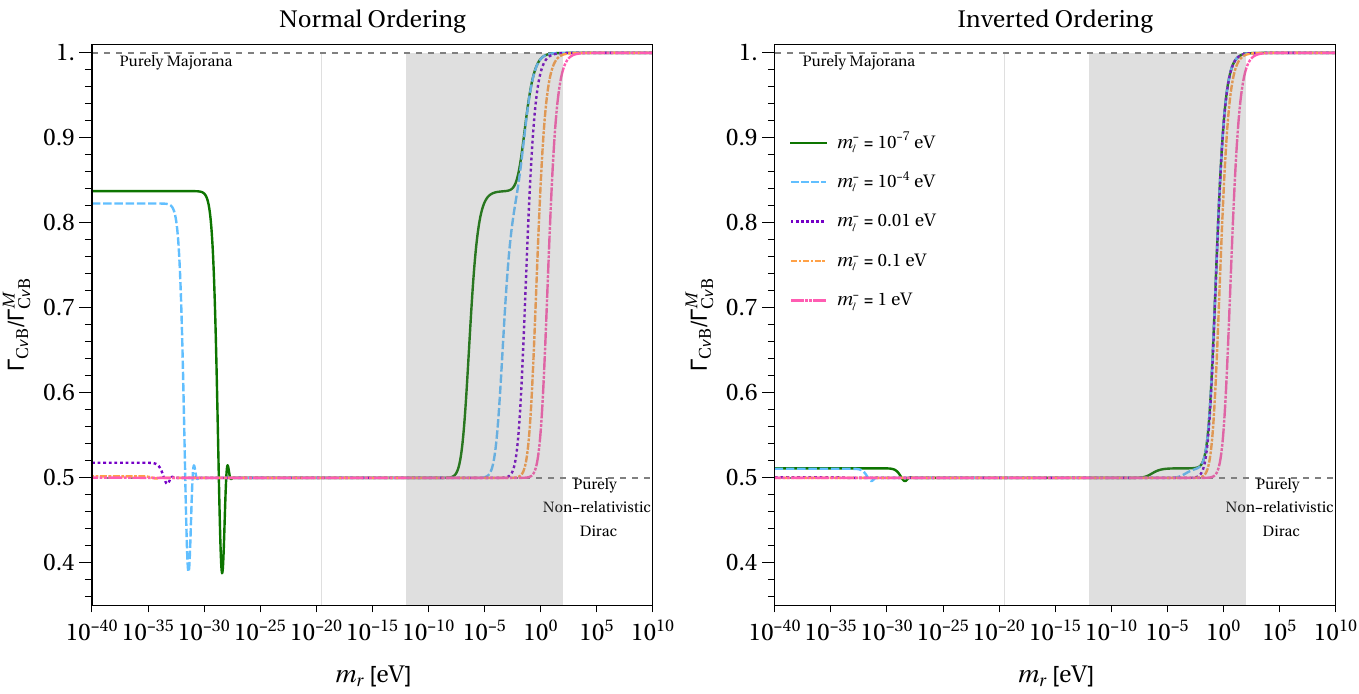}
\caption{Ratio of the $\CNB$ capture for the general Dirac+Majorana scenario to the purely Majorana rate, as a function of the scale of lepton number violation $m_r$ for lightest neutrino masses of $m_\ell^- = 10^{-7}$ eV (green), $m_\ell^- = 10^{-4}$ eV (light blue dashed), $m_\ell^- = 0.01$ eV (magenta dotted), $m_\ell^- = 0.1$ eV (orange dot-dashed), $m_\ell^- = 1$ eV (pink dot-dot-dashed) for the Normal (left) and Inverted (right) Orderings. The shaded regions are excluded from neutrino oscillation experiments. Note that for a relativistic lightest neutrino in the purely Dirac case,  the ratio tends to the value in Eq.~\eqref{eq:limDCR}.}
\label{fig:Rat}
\end{figure*}
Let us analyse the different limits in the capture rate Eq.~\eqref{eq:CR}. In the see-saw limit previously mentioned, where the mixing angle $\theta_{i}\to 0$, we have
\begin{align}\label{eq:CRPD}
    \Gamma_{\CNB} \approx 2N_T\overline{\sigma} n_0,
\end{align}
corresponding to the usual Majorana capture rate. 
Now if the mixing angle is maximal, $\cos\theta_{i}=\sin\theta_{i}= 1/\sqrt{2}$ and the fields $\nu_i^\pm$ are degenerate in mass, i.e.~$m_{r_i}=0$, the capture rate is
\begin{align}
    \Gamma_{\CNB} \approx N_T\overline{\sigma} n_0(1 + \sum_{i=1}^3 |U_{ei}|^2\langle v_i\rangle),
\end{align}
which is the value obtained for Dirac neutrinos~\cite{Roulet:2018fyh}.

Let us examine the ratio between the full neutrino capture rate and the purely Majorana case,
\begin{align}
\frac{\Gamma_{\CNB}}{\Gamma_{\CNB}^M} = 1+\sum_{i=1}^3|U_{ei}|^2\left[\cos^2\theta_{i} + \sin^22\theta_{i}\,\langle v_i\cos\left(\Delta \Phi_i\right)\rangle \right],
\end{align}
In this analysis, we assume that the values of $m_i^-$ coincide with the mass of the active neutrinos in the seesaw limit. 
Additionally, we consider all singlet masses to be equal, $m_{r_1}=m_{r_2}=m_{r_3}=m_r$.
In Fig.~\ref{fig:Rat}, we illustrate the behaviour of the ratio as it varies with the $m_r$ while maintaining a fixed value for $m_\ell^-$, the mass of the lightest state. 
We consider different fixed values for the lightest neutrino $m_\ell^- = 10^{-7}$ eV (green), $m_\ell^- = 10^{-4}$ eV (light blue dashed), $m_\ell^- = 0.01$ eV (magenta dotted), $m_\ell^- = 0.1$ eV (orange dot-dashed), $m_\ell^- = 1$ eV (pink dot-dot-dashed), for both the Normal (left) and Inverted (right) Orderings.

The shaded region indicates values that are excluded based on current neutrino oscillation experiments~\cite{deGouvea:2009fp}. 
As anticipated from the limits discussed earlier, particularly when the lightest neutrino is non-relativistic today, we observe that for $m_r \gg m_1^-$, the capture rate aligns with the purely Majorana scenario. 
Conversely, in the opposite limit, we recover the expected Dirac behaviour, consistent with the findings in the previously described pseudo-Dirac limit.
The transition between these two limits hinges on the mass spectrum of $m_i^\pm$. 
Specifically, as $m_r$ approaches approximately $0.1 m_i^-$, the mixing angle begins to deviate from maximal, resulting in an increased capture rate. 
When $m_r$ surpasses $m_i^-$ by roughly two orders of magnitude, the capture rate tends to approach the maximal value associated with the purely Majorana case.
However, it is important to highlight that the transition region, which could potentially yield varying capture rates, falls within the range excluded by current experimental data.

Significant differences arise when considering the scenario where the lightest neutrino remains relativistic in the present day. In this case, it is expected that the capture rate for Majorana neutrinos remains the same, while that for Dirac neutrinos increases, depending on the velocity and the PMNS mixing matrix element corresponding to the lightest neutrino.
For the normal ordering, the ratio takes on a value of approximately
\begin{align}\label{eq:limDCR}
\frac{\Gamma_{\CNB}^D}{\Gamma_{\CNB}^M} &\approx \frac{1}{2}(1 + |U_{e1}|^2\langle v_1\rangle)\approx 0.84
\end{align}
when the lightest neutrino is massless.
This is consistent with previous results in Ref.~\cite{Roulet:2018fyh}.

In our case, taking the case of $m_\ell^- = 10^{-7}$ eV, we expect $\Gamma_{\CNB}/\Gamma_{\CNB}^M\simeq 0.84$ for $m_r \ll 10^{-31}$ eV. This mimics the result expected for Dirac neutrinos. However, as $m_r$ increases, a distinctive pattern emerges. A minimum becomes apparent in the capture rate ratio.
This diminution in the ratio occurs due active-sterile transitions, which reach the first oscillation maximum when $\Delta\Phi_i = \pi$. 
In the relativistic lightest neutrino regime, we have that
\begin{align}
    \Delta\Phi_i=\frac{\delta m_\ell^2}{2 p} L_{\CNB},
\end{align}
where $\delta m_\ell^2 = (m_\ell^+)^2-(m_\ell^-)^2$, and $L_{\CNB}$ denotes the ${\CNB}$ propagation distance~\cite{Esmaili:2012ac,DeGouvea:2020ang}
\begin{align}
    L_{\CNB} &= \int_0^z \frac{dz^\prime}{(1+z^\prime)H(z^\prime)}\approx 2.35~{\rm Gpc},
\end{align}
for $z=10^{10}$ redshift value at neutrino decoupling. Thus, the oscillation maximum occurs when
\begin{align}
    \delta m_\ell^2 &= \frac{2 \pi p}{L_{\CNB}}\notag\\
    &\sim 10^{-35}~{\rm eV^2}\left(\frac{2.35~{\rm Gpc}}{L_{\CNB}}\right)\left(\frac{p}{0.6~{\rm meV}}\right)
\end{align}
Since $\delta m_\ell^2 = m_r(m_r+2m_\ell^-)\ll (m_\ell^-)^2$, we obtain the value of $m_r$ where the maximum active-sterile oscillation takes place, at approximately
\begin{align}
    m_r^{\rm osc}&\approx \frac{\pi p}{m_\ell^- L_{\CNB}}\notag\\
    &\sim 5\times 10^{-30}~{\rm eV} \left(\frac{1~{\rm \mu eV}}{m_\ell^-}\right)\left(\frac{2.35~{\rm Gpc}}{L_{\CNB}}\right)\left(\frac{p}{0.6~{\rm meV}}\right).
\end{align}
The averaging effect remains until $m_r$ surpasses a certain value, in this case equalling $10^{-10}$ eV in the normal ordering.

As $m_r$ is further increased, the growth rate increases again and makes a transition when $m_r\simeq  \sqrt{\Delta m_{\rm sol}^2} =8.7$ meV. This explains the second step-like feature in the plot. For more massive neutrinos, maximal mixing is preserved and we recover the Majorana capture rate. This behaviour is contingent upon the lightest neutrino mass $m_\ell^-$, and becomes less prominent as it increases.

Upon comparing the outcomes for both normal and inverted orderings, a notable distinction emerges concerning the capture rate for extremely small values of $m_r$.
In the case of the inverted ordering, where the lightest neutrino corresponds to $m_3^-$, its capture is governed by the small mixing angle $\theta_{13}$. 
Consequently, the asymptotic value for $m_r\ll m_r^{\rm osc}$ exhibits only a marginal correction of approximately $\sim 2.5\%$ from the non-relativistic Dirac scenario.

In summary, the overall behaviour of the capture rate critically hinges on the value of $m_r$.
When $m_r = m_r^{\rm osc}$, a minimum arises due to the active neutrinos undergoing a transition to sterile neutrinos. 
For values larger than $m_r^{\rm osc}$, maximal mixing prevails, and the active-sterile oscillation averages out, resulting in a capture rate akin to that of the purely Dirac case, until $m_r$ approaches the vicinity of $m_\ell^-$, where the mixing deviates from maximality, leading to a capture rate approaching the Majorana value.
On the other hand, for values lower than $m_r^{\rm osc}$, the capture rate tends toward the Dirac case, but with a correction due to the presence of a relativistic lightest neutrino.
Indeed, even when dealing with a relativistic lightest neutrino, the capture rate has the potential to align with the Dirac case in the non-relativistic limit. 
This phenomenon arises due to the active-sterile oscillations averaging out, leading to a cancellation between helicity contributions that effectively nullify the impact of having a relativistic lightest neutrino. 
Hence, the capture rate can converge to a value comparable to that in the Dirac case, despite the relativistic nature of the lightest neutrino.

\vspace{0.2in}
\section{Event Rates in a PTOLEMY-like Detector}
\label{sec:Ptolemy}

\begin{figure}[!t]
\centering
\includegraphics[width=0.95\linewidth]{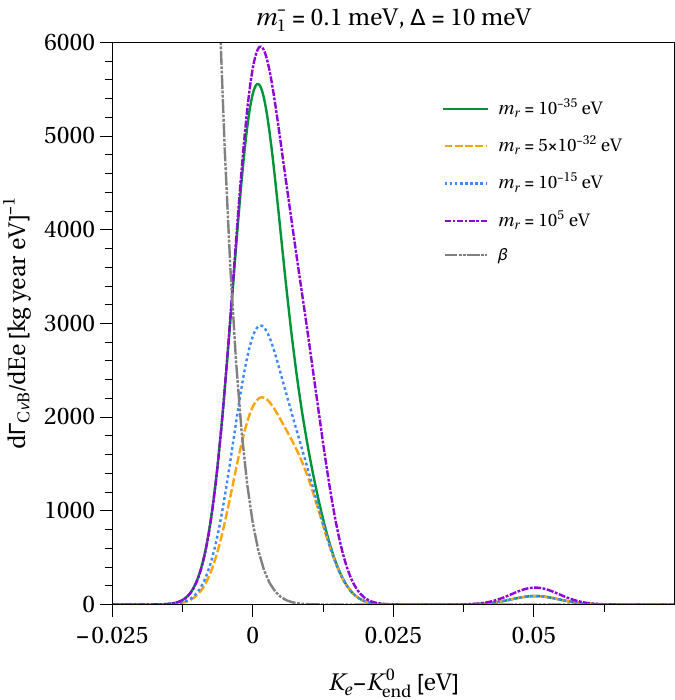}
\caption{Expected electron spectra as a function of the observed energy for different values of the singlet mass $m_r=10^{-35}$ eV (green), $m_r=5\times 10^{-32}$ eV (orange dashed), $m_r=10^{-15}$ eV (blue dotted), and $m_r=10^{5}$ eV (purple dot-dashed), assuming the normal ordering. We consider an experimental resolution with full width at half maximum $\Delta = 10$ meV, and the lightest neutrino mass of $m_1^- = 0.1$ meV. The $\beta$-decay background is plotted as a grey dot-dot-dashed line.}
\label{fig:evs}
\end{figure}

The proposed PTOLEMY experiment aims to detect neutrinos from the $\CNB$ utilizing a layer of graphene with atomic tritium on top of it~\cite{PTOLEMY:2018jst,PTOLEMY:2022ldz}. 
Although various setups for PTOLEMY have been considered, our focus lies in examining how the presence of singlets would impact the detection events in PTOLEMY or similar experiments. 
In the capture process described earlier, when a neutrino interacts with the tritium nucleus, it produces an electron whose energy can be measured using specific techniques. 
The kinematics of this capture process results in definite energy for the electrons~\cite{Long:2014zva}
\begin{align}\label{eq:energy_peak}
	E_e^{{\CNB}, i}\simeq m_e+K_{\rm end}^0+2\,m_i.
\end{align}
Here, $K_{\rm end}^0$ represents the endpoint energy of the electrons emitted from the $\beta$-decay of tritium. 
Given that the electrons produced after neutrino capture are monochromatic, they will generate one or more peaks at energies larger than $K_{\rm end}^0$.
The distinguishability of the $\CNB$ emitted electrons from those originating from tritium $\beta$-decay relies on the energy resolution. 
With a sufficiently high resolution, it becomes possible to differentiate these events. 
However, if the energy resolution is too large, the $\CNB$ electron events may be buried under a significant background.
To account for this, we convolve the capture in Eq~\eqref{eq:CRPD} with an assumed Gaussian-like experimental resolution,
\begin{widetext}
\begin{subequations}
	\begin{align}
	\frac{d\Gamma_{\CNB}}{dE_e}  &= \frac{1}{\sqrt{2\pi\sigma^2}}\sum_{j=1}^{3}\int_{-\infty}^\infty\, dE_e^\prime \,\Gamma_{\CNB}^j \,\exp\left[-\frac{(E_e^\prime-E_e)^2}{2\sigma^2}\right]\,\delta(E_e-E_e^{{\CNB}, j}), \\
	\frac{d\Gamma_{\beta}}{dE_e}&= \frac{1}{\sqrt{2\pi\sigma^2}}\int_{-\infty}^\infty\, dE_e^\prime \, \frac{d\Gamma_{\beta}}{dE_e^\prime}\,\exp\left[-\frac{(E_e^\prime-E_e)^2}{2\sigma^2}\right],
\end{align}
\end{subequations} 
\end{widetext}
where $\sigma$ is the energy resolution, also parameterised through the full width at half maximum (FWHM) $\Delta = 2.35\sigma$, and $\Gamma_{\CNB}^j$ indicates the capture rate associated with the $i$-th mass eigenstate. 
By utilizing the complete expression for the $\beta$-decay spectrum of tritium~\cite{Ludl:2016ane}, we present in Fig.~\eqref{fig:evs} the anticipated electron spectra as a function of the measured energy for various values of the singlet mass $m_r=10^{-35}$ eV (green), $m_r=m_r^{\rm osc}=5\times 10^{-32}$ eV (orange dashed), $m_r=10^{-15}$ eV (blue dotted), and $m_r=10^{5}$ eV (purple dot-dashed), assuming the normal ordering. 
We consider an FWHM of $\Delta = 10$ meV, and the lightest neutrino mass of $m_1^- = 0.1$ meV. 
The $\beta$-decay background is denoted by the grey dot-dot-dashed line.
In all cases, the electron spectrum exhibits two primary peaks. 
The first peak, with a maximum at $K_e - K^0_{\rm end} = m_1^-$, corresponds to the superposition of capture rates for the lightest neutrinos. 
The second peak emerges around the mass of the heaviest neutrino, approximately $K_e - K^0_{\rm end} \approx 50$ meV.
Furthermore, the extreme values of $m_r=10^{-35}$ eV and $m_r=10^{5}$ eV depict the event spectra for Dirac, encompassing a relativistic lightest neutrino, and Majorana, respectively.
A significant difference between these two cases appears due to the contribution of the heaviest neutrinos, which change the shape of the first peak, and enhance the capture of the heaviest states.
Meanwhile, for the intermediate value of $m_r=10^{-15}$ eV, the capture rate has a value corresponding to the Dirac case for a non-relativistic spectrum.
Regarding the $m_r=m_r^{\rm osc}$ scenario, we observe a reduction in the spectrum, even when compared to the Dirac case. 
As previously discussed, this reduction stems from the oscillation of active neutrino states into sterile, and thus unobservable, neutrinos, thereby reducing the number of states available for capture. 
Additionally, the electron spectrum is no longer symmetric in this instance due to the emergence of the peak associated with the capture of the superposition $\nu_2^\pm$.
These findings underscore the vital role of underlying mass generation in neutrino capture, particularly in a PTOLEMY-like experiment.

\section{Conclusions}
\label{sec:conc}
Possible future detection of the cosmic neutrino background will be a watershed moment in our understanding of the early Universe, as well as the nature of the neutrinos. In particular, it is expected to shed light on whether neutrinos are Dirac or Majorana, thereby offering a probe of lepton number violation in our Universe. Currently, the most popular idea for the detection of the \cnub involves neutrino capture on tritium - an idea which is being actively pursued by the PTOLEMY collaboration.

In this paper, we studied the dependence of the neutrino capture on the extent of lepton number violation in the Standard Model. We focused on pseudo-Dirac neutrinos, where lepton number can be softly broken so that neutrinos behave as Dirac while actually being Majorana. In such a scenario, we showed that the neutrino capture rate smoothly transitions between a purely Dirac case and a purely Majorana case. As a result, even a slight deviation of the capture rate from the purely Dirac case can signal a soft violation of lepton number.

Active-sterile oscillations, mediated by a tiny mass-squared difference, can also cause a distortion in the capture rate. We found that in the scenario where the lightest neutrino is relativistic, the distortion can be sensitive to the value of the mass-squared difference as small as $\dm\sim 10^{-35}~\eV^2$. From this value, and depending on the mass of the lightest neutrino, there exists a critical Majorana mass scale, $m_r^{\rm osc}$, such that for $m_r \ll m_r^{\rm osc}$, the capture rate approaches the Dirac rate, but with an enhancement due to the presence of the relativistic lightest neutrino. As $m_r$ approaches $m_r^{\rm osc}$, active-sterile oscillations take over leading to an overall minima in the capture rate, which can go below the Dirac rate as well. On the other hand, for $m \gg m_r^{\rm osc}$, active-sterile oscillations average out, and the Dirac rate is recovered. This happens until the $m_r$ approaches the value of the lightest neutrino, where the active-sterile mixing gradually deviates from the maximum, and the capture rate approaches the Majorana value.

We compared the neutrino capture rates in a PTOLEMY-like detector as a function of the sterile neutrino mass - which is a measure of the strength of lepton number violation. We confirmed that a detector like PTOLEMY would indeed be sensitive to the underlying mechanism of neutrino mass generation. The electron spectra events are shown to lie between a purely Dirac hypothesis and a purely Majorana hypothesis, with the exact rate depending on the value of the sterile neutrino mass. 

Through this analysis, we pointed out the sensitivity of the capture rate of the \cnub to the mechanism connecting neutrino mass-generation. We performed a simple analysis under the approximation where the underlying $6\times6$ neutrino mixing matrix, consisting of 3 active and 3 sterile neutrinos, factorizes into 3 independent  $2\times2$ matrix involving active-sterile pairs. Future studies will be aimed at relaxing this approximation to test the sensitivity of our results on the underlying neutrino mixing mechanism.

\section*{Acknowledgments}

We would like to thank André de Gouvea for helpful discussions in the initial stages of the project, and for the insightful comments on the first version of this manuscript.
YFPG would like to thank the warm hospitality of the Particle and Astroparticle Division of the Max-Planck-Institute f\"ur Kernphysik where part of this work was completed.
This work has been funded by the UK Science and Technology Facilities Council (STFC) under grant ST/T001011/1. 
This project has received funding/support from the European Union’s Horizon 2020 research and innovation programme under the Marie Sk\l{}odowska-Curie grant agreement No 860881-HIDDeN.
This work has made use of the Hamilton HPC Service of Durham University.

\appendix


\bibliographystyle{apsrev4-1}
\bibliography{refs.bib}

\end{document}